\documentclass[epj]{svjour}
\input epsf

\newcounter{append}
\newcommand{\bc}{\begin{center}}
\newcommand{\ec}{\end{center}}
\newcommand{\be}{\begin{equation}}
\newcommand{\ee}{\end{equation}}
\newcommand{\ba}{\begin{array}}
\newcommand{\ea}{\end{array}}
\newcommand{\beqn}{\begin{eqnarray}}
\newcommand{\eeqn}{\end{eqnarray}}

\begin{document}

\title{Scaling behaviour of the relaxation in quantum chains}

\author{Dragi Karevski }
\institute{Laboratoire de Physique des Mat\'eriaux, Universit\'e Henri
Poincar\'e (Nancy 1), B.P. 239,\\ F-54506 Vand\oe uvre l\`es Nancy cedex,
France}

\date{November 21, 2001}

\abstract
{We consider the  nonequilibrium time evolution of the transverse magnetization in the critical Ising 
and $XX$ quantum chains. For some inhomogeneously magnetized initial states we derive analytically 
the transverse magnetization profiles and show that they evolve into scaling forms in the long-time limit.
In particular it is seen that the Ising chain exhibits some similarities with the conserved dynamics $XX$ chain.
That is, after a transient regime, the total residual magnetization in the transverse direction is also conserved in the 	
Ising case. A class of general initial states is also considered.
}

\authorrunning{D. Karevski }
\titlerunning{Relaxation and scaling functions in quantum chains}


\maketitle

\maketitle

\section{Introduction}
Nonequilibrium properties of quantum systems have attracted a lot of interest since they have natural 
dynamics in contrast to classical ones and since classical effects are actually quantal. There are several 
ways to consider nonequilibrium quantum systems. One is to couple the quantum mechanical system to a 
heat bath which can be itself described quantum mechanically~\cite{weiss}. In this case, a part of the whole
system is isolated and called `the system' while the rest is supposed to describe a certain environnement
with which the system interacts and dissipates through. Another route is to impose a current on the system
and investigate the steady states~\cite{racz1}. Still an other possibility is simply to investigate the relaxation
of an initial state, in which the system has been prepared, and evaluate expectation values of observables
at later times. This was done recently~\cite{racz2} on the $XX$-quantum chain with a step-like magnetization 
initial state.  More recently in ref.~\cite{berim}, the relaxation of spatially inhomogeneous initial states has been treated 
for several variants of the XY quantum model.
Relaxation phenomena at zero temperature with homogeneous
initial state has been considered in ref.~\cite{trimper},\cite{igloi} for the $XX$ and Ising chains in a transverse field
in the context of aging. 

In this work we study the nonequilibrium profiles of the critical Ising and $XX$ quantum chains.
We suppose that at the initial time $t=0$ the system is prepared in a given state $|\Psi\rangle$.
The time evolution of the system is entirely governed by the Schr\"odinger equation and is formally given by
\begin{equation}
|\Psi(t)\rangle= \exp(-iHt)|\Psi\rangle
\end{equation}
since the systems under consideration are closed.
The basic quantity we calculate is the expectation value of
the local transverse magnetization at time $t$, 
$m(l,t)\equiv\langle\Psi(t)|\sigma_{l}^{z}|\Psi(t)\rangle$.
We first consider two different initial states $|\Psi\rangle$,
a kink in the $z$~direction, $|\dots\uparrow\uparrow\uparrow\downarrow\downarrow\downarrow\dots\rangle$, 
which was already considered in ref.~\cite{racz2} for the quantum $XX$~chain, and a droplet configuration 
$|\dots\uparrow\uparrow\downarrow..\downarrow\uparrow\uparrow\dots\rangle$ for both Ising and $XX$~chains. 
In ref.~\cite{trimper} the cumulated magnetization was calculated for the $XX$~chain but the profile
itself was not considered. We give finally the general expression for the relaxation of the transverse magnetization 
in terms of a convolution product in the continuum limit. The kernels of both XX and Ising chains are readly expressed
in both direct and Fourier space.

\section{Basic Quantities}
The one dimensional Ising and $XX$ Hamiltonians with $L$ sites and open boundary conditions are given by the same one-parameter
anisotropic $XY$ Hamiltonian:
\begin{equation}
H=-\frac{1}{2}\sum_{k=1}^{L-1}\left[\frac{1+\kappa}{2}\sigma_{k}^{x}\sigma_{k+1}^{x}+
\frac{1-\kappa}{2}\sigma_{k}^{y}\sigma_{k+1}^{y}\right] -\frac{h}{2}\sum_{k=1}^{L}\sigma_{k}^{z}
\label{eq1}
\end{equation}
where the anisotropy parameter $\kappa=1$ corresponds to the Ising case with a $Z_{2}$ symmetry 
and $\kappa=0$ describes the  $XX$-model which has $U(1)$ symmetry. 
The Hamiltonian~(\ref{eq1}) is diagonalizable through of a Jordan-Wigner
transformation, followed by a canonical transformation~\cite{lieb}. In terms of the 
Clifford operators $\{\Gamma_{l}^{i}\}$, the Jordan-Wigner transformation is expressed as~\cite{hinri}
\begin{equation}
\begin{array}{l}
\Gamma_{n}^{1}=(-1)^{n-1}\left(\prod_{j=1}^{n-1}\sigma^{z}_{j}\right) \sigma^{x}_{n} \\
\Gamma_{n}^{2}=-(-1)^{n-1}\left(\prod_{j=1}^{n-1}\sigma^{z}_{j}\right) \sigma^{y}_{n}\; .
\end{array}\label{jw}
\end{equation}
The generators $\{\Gamma_{l}^{i}\}$ satisfy 
\begin{equation}
\langle\Gamma_{n}^{i}|\Gamma_{k}^{j}\rangle=\delta_{ij}\delta_{nk}\; , \qquad (n,l=1,\dots,L;\; i,j=1,2)\; ,
\end{equation}
where we have introduced a pseudoscalar product defined as $\langle C|D\rangle\equiv \frac{1}{2}\{C,D\}$ with
$\{.,.\}$ the anticommutator. The original spin variables are obtained in terms of the $\Gamma$s
by inverting the previous relations. One obtains $\sigma^{x}_{n}\sigma^{x}_{n+1}=-i\Gamma^{2}_{n}\Gamma^{1}_{n+1}$, 
$\sigma^{y}_{n}\sigma^{y}_{n+1}=-i\Gamma^{2}_{n+1}\Gamma^{1}_{n}$ and $\sigma^{z}_{n}=-i\Gamma^{2}_{n}\Gamma^{1}_{n}$, 
so that~(\ref{eq1}) is written as
\begin{equation}
H=\frac{i}{4}\sum_{k}-\Gamma_{k}^{\dagger}(i\sigma^{y})\Gamma_{k+1}+\kappa\Gamma_{k}^{\dagger}\sigma^{x}\Gamma_{k+1}
-h\Gamma_{k}^{\dagger}i\sigma^{y}\Gamma_{k}\; ,\label{Hgam}
\end{equation}
where ${ {\Gamma}_{k}^{\dagger}}=({\Gamma^{1}_{k}},{\Gamma^{2}_{k}})$, the hermitian conjuguate of $\Gamma_{k}$,
is a $2$-components spinor and $\sigma^{x}$ and $\sigma^{y}$ are the Pauli matrices.
Introducing the $2L$-components Clifford operator ${\bf \Gamma}^{\dagger}=(\Gamma_{1}^{\dagger},\Gamma_{2}^{\dagger},\dots,\Gamma_{L}^{\dagger})$,
we arrive at $H=(1/4){\bf \Gamma}^{\dagger}{\bf T}{\bf \Gamma}$ with ${\bf T}^{\dagger}={\bf T}$.
The diagonalisation is then performed by the introduction of the diagonal Clifford generators 
$\gamma_{q}^{\dagger}=(\gamma^{1}_{q},\gamma^{2}_{q})$
related to the old one by
$\Gamma^{1}_{l}=\sum_{q}\phi_{q}(l)\gamma^{1}_{q}$ and 
$\Gamma^{2}_{l}=\sum_{q}\psi_{q}(l)\gamma^{2}_{q}$ with real $\phi$ and $\psi$ components. Introducing the Fermi operators 
$\eta_{q}=1/2(\gamma^{1}_{q}+i\gamma^{2}_{q})$ and the adjoint $\eta_{q}^{\dagger}$, finally one arrives at the usual free fermionic Hamiltonian
$H=\sum_{q}\epsilon_{q}\eta_{q}^{\dagger}\eta_{q}+E_{0}$.
The excitations energies $\epsilon_{q}$ and the transformation coefficients $\phi_{q}(l),\psi_{q}(l)$ are solution 
of the $2L\times 2L$ eigenvalue system ${\bf T}{\bf V}_{q}=\epsilon_{q}{\bf V}_{q}$, with  components
$V_{q}^{\dagger}(k)=(\phi_{q}(k),-i\psi_{q}(k))$. The eigenvectors satisfy the orthogonality relations $\sum_{q}\phi_{q}(i)\phi_{q}(j)=\delta_{ij}$
and $\sum_{q}\psi_{q}(i)\psi_{q}(j)=\delta_{ij}$.

The time evolution of the spin operators are easily expressed in terms of the time dependence of the Clifford generators
${\Gamma}$ (see Appendix). The basic time evolution of the diagonal operators is
$\gamma_{q}(t)=e^{iHt}\gamma_{q}e^{-iHt}={\cal R}(\epsilon_{q}t)\gamma_{q}$ where
${\cal R}(\theta)$ is a rotation of angle $\theta$. In matrix form we have 
\begin{equation}
\left(
\begin{array}{l}
\gamma^{1}_{q}(t)\\
\gamma^{2}_{q}(t)\end{array}\right) =\left(
\begin{array}{cc}
\cos\epsilon_{q}t  & \sin\epsilon_{q}t\\
-\sin\epsilon_{q}t  & \cos\epsilon_{q}t\end{array}\right)\left(\begin{array}{l}
\gamma^{1}_{q}\\
\gamma^{2}_{q}\end{array}\right)\; .
\label{rot}
\end{equation}
Using this relations, we can express the time dependence of the ${\Gamma}$s through an expansion onto 
the basis $\{\Gamma_{k}^{i}\}$:
\begin{equation}
\Gamma_{n}^{j}(t)=e^{iHt}\Gamma_{n}^{j}e^{-iHt}=
\sum_{k,i} \langle\Gamma_{k}^{i}|\Gamma_{n}^{j}(t)\rangle\; \Gamma_{k}^{i} 
\label{exp1}
\end{equation}
with components
\begin{eqnarray}
\langle \Gamma^{1}_{k}|\Gamma^{1}_{l}(t)\rangle&=&\sum_{q}\phi_{q}(k)\phi_{q}(l)\cos \epsilon_{q}t \nonumber\\
\langle \Gamma^{1}_{k}|\Gamma^{2}_{l}(t)\rangle&=&\langle \Gamma^{2}_{l}|\Gamma^{1}_{k}(-t)\rangle
=-\sum_{q}\phi_{q}(k)\psi_{q}(l)\sin \epsilon_{q}t\nonumber\\
\langle \Gamma^{2}_{k}|\Gamma^{2}_{l}(t)\rangle&=&\sum_{q}\psi_{q}(k)\psi_{q}(l)\cos \epsilon_{q}t \; .
\label{comp}
\end{eqnarray}

For the Ising chain, at the critical point $h=1$, the basic contractions are
obtain in a closed form. For open boundary conditions, the excitation energies $\epsilon_{q}=2\sin(q/2)$ and eigenvectors 
$\phi$ and $\psi$ are~\cite{turban}: 
\begin{eqnarray}
\phi_{q}(l)&=&(-1)^{l}\frac{2}{\sqrt{2L+1}}\cos q(l-1/2)\nonumber\\
\psi_{q}(l)&=&(-1)^{l+1}\frac{2}{\sqrt{2L+1}}\sin ql
\end{eqnarray}
with $q=(2p+1)\pi/(2L+1)$. In the thermodynamic limit $L\rightarrow\infty$, the contractions are then expressed in terms of 
Bessel functions  $J_{n}(z)$ of integer order as~\cite{grad}:
\begin{eqnarray}
\langle \Gamma^{1}_{k}|\Gamma^{1}_{l}(t)\rangle& =&\langle \Gamma^{2}_{k}|\Gamma^{2}_{l}(t)\rangle =(-1)^{k+l} J_{2(l-k)}(2t)\nonumber\\
\langle \Gamma^{1}_{k}|\Gamma^{2}_{l}(t)\rangle &=&-(-1)^{k+l+1} J_{2(l-k)+1}(2t)\; .
\label{I1}
\end{eqnarray}

For the $XX$-chain in a similar way one obtains 
\begin{eqnarray}
\langle \Gamma^{1}_{k}|\Gamma^{1}_{l}(t)\rangle =(i)^{l-k} J_{l-k}(t)\left\{
\begin{array}{l}
\cos (ht)\; ;\;  l-k=2p\\
-i\sin (ht) \; ;\;  l-k=2p+1
\end{array}\right.\nonumber\\
\langle \Gamma^{1}_{k}|\Gamma^{2}_{l}(t)\rangle =(i)^{l-k} J_{l-k}(t)\left\{
\begin{array}{l}
-\sin (ht)\; ; \; l-k=2p\\
-i\cos (ht) \; ;\;  l-k=2p+1
\end{array}\right.
\label{XX1b}
\end{eqnarray}
and $\langle \Gamma^{2}_{k}|\Gamma^{2}_{l}(t)\rangle = \langle \Gamma^{1}_{k}|\Gamma^{1}_{l}(t)\rangle $.

In what follows we are interested in the time relaxation of the transverse magnetization. In the Heisenberg picture,
we have, using the  expansion~(\ref{exp1}),
$\sigma^{z}_{l}(t)=-i\Gamma^{2}_{l}(t)\Gamma^{1}_{l}(t)=-i\sum_{k1,k2}^{i1,i2}
\langle \Gamma_{k1}^{i1}|\Gamma^{2}_{l}(t)\rangle\langle \Gamma_{k2}^{i2}|\Gamma^{1}_{l}(t)\rangle\Gamma_{k1}^{i1}\Gamma_{k2}^{i2}$. The
expectation value in the z-direction $|\Psi\rangle$ state is then
\begin{eqnarray}
m(l,t)&=&\sum_{k}\Big[\langle \Gamma^{2}_{k}|\Gamma^{2}_{l}(t))\langle \Gamma^{1}_{k}|\Gamma^{1}_{l}(t)\rangle\nonumber\\
&-&\langle \Gamma^{1}_{k}|\Gamma^{2}_{l}(t)\rangle\langle \Gamma^{2}_{k}|\Gamma^{1}_{l}(t)\rangle\Big]\langle\Psi|\sigma_{k}^{z}|\Psi\rangle
\label{sz1}
\end{eqnarray}
since only the terms $\langle\Psi|\Gamma^{2}_{k}\Gamma^{1}_{k}|\Psi\rangle$ are non-vanishing in the state $|\Psi\rangle$.

\section{Kink-like initial state}
We consider first the initial state with a kink located at the origin
\begin{equation}
|\Psi\rangle=|\uparrow\rangle^{\otimes^{N/2}}\otimes|\downarrow\rangle^{\otimes^{N/2}}
\end{equation}
where $|\uparrow,\downarrow\rangle$ are the eigenstates of the $\sigma^{z}$ Pauli matrix,
$\sigma^{z}|\uparrow,\downarrow\rangle=\pm|\uparrow,\downarrow\rangle$, and we take the thermodynamic 
limit $N\rightarrow\infty$. We present here only the results for the critical ($h=1$) Ising quantum chain 
since the $XX$ chain
was already considered in ref.~\cite{racz2}.
Using the previous formulas~(\ref{sz1}) and~(\ref{I1}) for the contractions, a straightforward calculation leads to
\begin{equation}
m(l,t)=-\sum_{p=1-l}^{l-1} \left[\left(1-\left(\frac{p}{t}\right)^{2}\right)J_{2p}^{2}(2t)+
{J_{2p}^{'}}^{2}(2t)\right]\; .
\end{equation}
Now the analysis proceeds along the same lines as in ref.~\cite{racz2}. We introduce the discrete derivative
\begin{equation}
\begin{array}{rl}
\Phi_{n}^{'}(v) &\equiv  -t[m(n+1,t)-m(n,t)]_{n/t=v} \\
&=-2\frac{n}{v}[(1-v^{2})J_{2n}^{2}(2t)+{J_{2n}^{'}}^{2}(2t)]
\end{array}
\label{phi}
\end{equation}
For obvious symmetry reasons, we will consider only the part $v>0$ since 
we have $m(-n,t)=-m(n,t)$. Due to the different asymptotic properties of the Bessel functions one 
has to distinguish between the cases $v>1$ and $v<1$.
For $v>1$, which means that $n>t$, we are in the acausal region (outside the light-cone) since the 
excitations, traveling with velocity one~\cite{remarq}, have no time to propagate from
the initial position of the kink to the site $n$. Then it is the local environnement which completly governs
the behavior of the magnetization. That is, the magnetization relaxes as if the initial state was the completly
ordered state and one has 
\begin{equation}
m(n,t)=-\frac{1}{2}-\frac{1}{4t}J_{1}(4t)
\end{equation}
for $n>t$. 
The local magnetization reaches the constant value $-1/2$ with corrections of order $t^{-3/2}$ so that the derivative
$\Phi^{'}(v)$ is essentially vanishing for large $n$. This is exactly what is seen from the asymptotic behavior
of the Bessel functions and their derivatives, vanishing as $\exp[-\lambda(v) n]$ with $\lambda(v)>0$~\cite{grad}.

Inside the light-cone ($v<1$), with the help of the asymptotics for $\nu\gg 1$~\cite{grad}
\begin{equation}
\begin{array}{c}
J_{\nu}(\frac{\nu}{\cos\beta})=\sqrt{\frac{2}{\pi\nu\tan\beta}}\cos \psi\\
J^{'}_{\nu}(\frac{\nu}{\cos\beta})=-\sqrt{\frac{\sin2\beta}{\pi\nu}}\sin \psi
\end{array}
\end{equation}
where $\psi\equiv\nu(\tan\beta-\beta)-\pi/4$, one obtains for the derivative $\Phi^{'}_{n}(v)$:
\begin{equation}
\Phi^{'}_{n}(v)=-\frac{2}{\pi}\sqrt{1-v^{2}}=\Phi^{'}(v)
\label{sc1}
\end{equation}
which is $n$ independent due to the exact cancellation of the $\sin$ and $\cos$ terms in~(\ref{phi}).
Finally, by simple integration we obtain $m(n,t)=\Phi(n/t)$ with the scaling function
\begin{equation}
\Phi(v)=\left\{\begin{array}{lc}
1/2& v<-1\\
-\frac{1}{\pi}[v\sqrt{1-v^{2}}+\arcsin v]&\quad -1<v<1\\
-1/2& v>1
\end{array}\right.
\label{isingsc}
\end{equation}
This has to be compared with the $XX$ chain result~\cite{racz2}
$-\frac{2}{\pi}\arcsin v$ and $\pm 1$ outside the causal region. Contrary to the $XX$ chain
which has a conserved dynamics (the total z-component of the magnetization is a constant of motion)
in the Ising quantum chain there is a transient regime where the local magnetization relaxes faster than
$t^{-1}$ toward the stationary value $\pm 1/2$ and then only the residual kink spreads as in the $XX$ chain.
The results are shown in figure~1 where the inset describes the initial transient regime.
\begin{figure}[h]
\epsfxsize=8truecm
\begin{center}
\mbox{\epsfbox{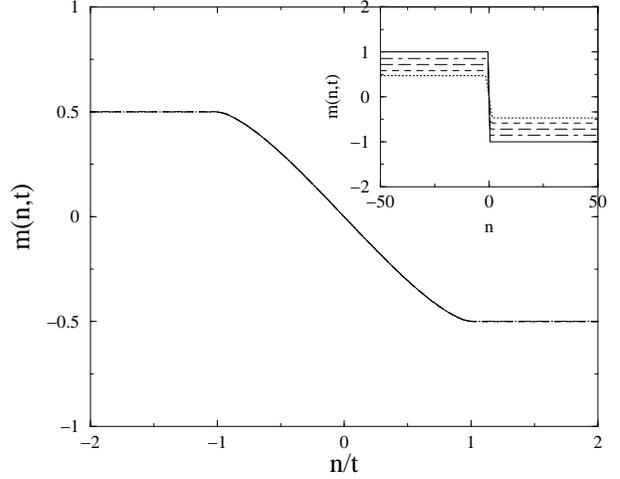}}
\end{center}
\caption{\label{fig1} Nonequilibrium transverse magnetization scaling function for the Ising quantum chain. The analytical 
expression~(\ref{isingsc}) and the numerical results are indistinguishable. In the inset, the transient regime is shown for times
smaller than $t=2$. The magnetization relaxes toward the value $\pm 1/2$.}
\end{figure}

\section{Droplet-like initial state}
Let us consider now the following initial state 
\begin{equation}
|\Psi\rangle=|\dots\uparrow \uparrow\uparrow\Downarrow \uparrow \uparrow\uparrow\dots\rangle
\label{droplet}
\end{equation}
with $|\Downarrow\rangle={|\downarrow\rangle^{\otimes}}^{L}$, that is a droplet of $L$ down spins
inside a bath of up spins, both interacting and evolving with the quantum Hamiltonian $H$.
This can be considered as a toy model for a quantum system (the midle part) coupled to some environnement
(the external part), both governed by the same microscopic interactions and one can study how the system part
relaxes due to the coupling to the external degrees of freedom.
We start with a one spin droplet within the Ising model.
In this case, using (\ref{sz1}) together with (\ref{droplet}), the Ising transverse magnetization is given by 
\begin{equation}
m(l,t)=\frac{1}{2}+\frac{1}{4t}J_{1}(4t)+\frac{1}{t}\Phi^{'}_{l}(l/t)
\end{equation}
where $\Phi^{'}_{l}(v)$ is the function introduced in the previous section.
For $v>1$ the magnetization is dominated by the first two terms since then the 
function $\Phi^{'}_{l}(l/t)$ is exponentially small. 
On the other hand for $v<1$, after the faster relaxation toward $1/2$ a scaling regime emerges
for the local excess magnetization, that is
\begin{equation}
m^{c}(l,t)\equiv m(l,t)-1/2= t^{-1}\Phi^{'}(\frac{l}{t})
\end{equation}
where $\Phi^{'}(v)$ is given by~(\ref{sc1}).

More generaly, for a droplet of size $L$, the transverse magnetization at time $t$
is given by
\begin{equation}
m(l,t)=\frac{1}{2}+\frac{1}{4t}J_{1}(4t)+\frac{1}{t}\sum_{k=-L/2}^{L/2}\Phi^{'}_{l-k}((l-k)/t)\; .
\end{equation}
In the light-cone the excess magnetization is given to the dominant order in $t^{-1}$ by
\begin{equation}
m^{c}(l,t)= t^{-1}\int_{-L/2}^{L/2}\Phi^{'}((l-k)/t) {\rm d}k=\int_{l/t-L/2t}^{l/t+L/2t}\Phi^{'}(u){\rm d}u
\end{equation}
so that finally in the scaling regime $l\gg L$ we have simply
\begin{equation}
m^{c}(l,t)=\frac{L}{t}\Phi^{'}(\frac{l}{t})=-\frac{2L}{\pi t}\sqrt{1-\left(\frac{l}{t}\right)^{2}}\; .
\end{equation}
For $l\sim {\cal O}(L)$, with $J_{n}(z)\sim\sqrt{2/\pi z} \cos(z-n\pi/4-\pi/4)$ for large $z$, the local 
magnetization excess is simply $-2L/\pi t$ plus subdominant corrections.
Then the total magnetization remaining inside the initial droplet region is 
\begin{equation}
M^{c}(t)=\int_{L} m^{c}(l,t){\rm d}l=-\frac{2L^{2}}{\pi t}
\end{equation}
which is exactly what was obtained in ref.~\cite{trimper} for the $XX$ quantum chain. As in the $XX$ case, where the total 
magnetization is conserved, something similar happens in the Ising case. If one considers the total
magnetization excess at time $t$, given by the integral over the whole space, we simply have a constant: 
\begin{equation}
\int_{-\infty}^{\infty} m^{c}(l,t){\rm d}l\simeq L\int_{-1}^{1} \Phi^{'}(v){\rm d}v=-L
\end{equation}
where $-L$ is the residual excess magnetization after the initial transient regime since the up(down) domain relaxes
locally toward $1/2$($-1/2$). This means that after the initial loss of magnetization, which takes place on microscopic time scales
of order $t\simeq 1/h=1$, the dynamic is conservative.
We have a conservative deviation around the stationary value. In fact, one can show that the local magnetization excess satisfy
a lattice continuity equation $\partial_{t}m^{c}(l,t)+j(l,t)-j(l-1,t)=0$, with the current density $j(l,t)$ given by
\begin{eqnarray}
j(l,t)=-2\sum_{k=-L/2}^{L/2} J_{2(l-k)-1}(2t)J_{2(l-k)}(2t)\nonumber\\
-J_{2(l-k)-2}(2t)J_{2(l-k)+1}(2t)
\end{eqnarray}
which is related to the expectation value of $\Gamma^{2}_{l-1}\Gamma^{2}_{l}\propto \sigma_{l-1}^{x}\sigma_{l}^{y}$. In the continuum limit, 
the current is simply expressed as $j(x,t)=\frac{x}{t}m^{c}(x,t)$.

Although, the total residual magnetization in the system part was calculated in ref.~\cite{trimper} for the $XX$ chain, 
the scaling profile was not considered. 
Using equations~(\ref{XX1b}) and~(\ref{sz1}) we have for all values of the transverse field $h$ 
\begin{equation}
m^{c}(l,t)\equiv m(l,t) -1=-2\sum_{k=-L/2}^{L/2}J^{2}_{l-k}(t)\; .
\end{equation}
In the light-cone, together with the asymptotic expressions for the Bessel functions, we obtain in the scaling regime
$t>l\gg L$
\begin{equation}
m^{c}(l,t)=\frac{L}{t}\Psi\left(\frac{l}{t}\right)
\end{equation}
with the scaling function
\begin{equation}
\Psi(v)=-\frac{2}{\pi}\frac{1}{\sqrt{1-v^{2}}}\; .
\end{equation}
One may verify that the integral over the whole space of the local magnetization gives back $-2L$ as it should be
for the conserved dynamics system under consideration.
The scaling functions for both Ising and $XX$ chains are presented in figure~2, where the analytical results
are compared with numerics.
\begin{figure}[h]
\epsfxsize=8truecm
\begin{center}
\mbox{\epsfbox{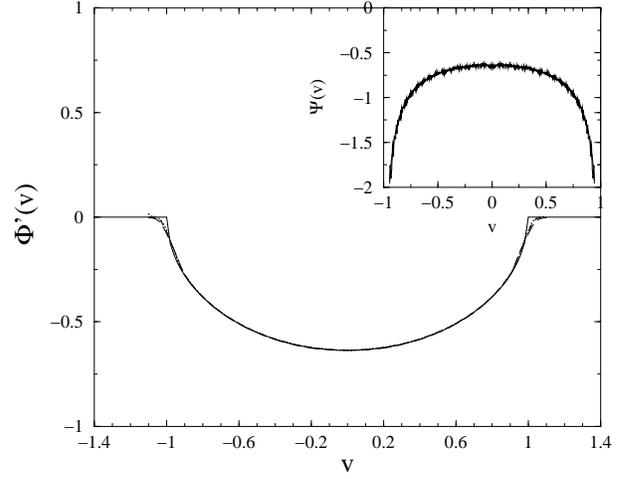}}
\end{center}
\caption{\label{fig2} Scaling functions for the critical Ising and $XX$(inset) quantum chains. 
The oscillations in the $XX$ case are finite size effects due to the initial droplet size.
One can see that as the time is increased the numerical results are closer and closer to the 
analytical scaling function. This can be seen more evidently at the boundaries $v=\pm 1$ for 
the Ising chain.}
\end{figure}
\section{General initial $z$-state}
Clearly, for translation invariant Hamiltonians, equation~(\ref{sz1}) giving the transverse magnetization is 
a discret convolution product:
\begin{equation}
m(l,t)=\sum_{k=-\infty}^{\infty}F_{t}(l-k) S(k)=(F_{t} * S)(l)
\label{conv1}
\end{equation}
with $S(k)=\langle\Psi|\sigma^{z}_{k}|\Psi\rangle$. The kernel $F_{t}(l)$ is given in the continuum limit by
$F_{t}(l)=\frac{1}{t} f(\frac{l}{t})$ with
\begin{equation}
f_{\kappa}(v)=\left\{\begin{array}{ll}
\frac{1}{\pi}\left(1-v^{2}\right)^{\kappa-1/2}& |v|<1\\
0& |v|>1
\end{array}\right.
\end{equation}
where the $\kappa=1$ refers to the Ising case and  $\kappa=0$ to the $XX$ chain.
The local magnetization $m(x,t)=m_{t}(v)$, is then expressed as the convolution product
\begin{equation}
m_{t}(v)=(S_{t} * f)(v)
\end{equation}
with $S_{t}(v)=S(tv)$. For the kink like initial state, we have $S_{t}(v)=-sgn(v)=1-2H(v)$, where $H(v)$ is the Heaviside
function. For the droplet-like initial state, 
\begin{equation}
S_{t}(v)=1-\frac{2L}{t}\left(\frac{t}{L}\right)\Pi(\frac{tv}{L})\; , 
\end{equation}
where
$\Pi(x)$ is the caracteristic function of the interval $[-1/2,1/2]$. In the long time limit, $t\gg L$, we have
$S_{t}(v)=1-\frac{2L}{t}\delta(v)$, so that we recover very simply the results of the previous section.

For a general initial state in the z-direction, 
the Fourier transform of equation~(\ref{conv1}) is 
\begin{equation}
\tilde{m}_{t}(q)=\tilde{S}_{t}(q)  \tilde{f}(q)
\end{equation}
with
\begin{equation}
\tilde{f}_{\kappa}(q)=
\frac{1}{\pi}\int_{-1}^{1}(1-v^{2})^{\kappa-1/2}e^{-2i\pi qv}dv
\end{equation}
so that the kernels in Fourier space are simply given by
\begin{equation}
\tilde{f}_{0}(q)=J_{0}(2\pi q)
\end{equation}
for the XX chain and
\begin{equation}
\tilde{f}_{1}(q)=\frac{J_{1}(2\pi q)}{2\pi q}=\frac{1}{2}\big(J_{0}(2\pi q)+J_{2}(2\pi q)\big)\; 
\end{equation}
for the Ising chain.
By inverse Fourier transform, it is possible to obtain the desired magnetization profile in direct space.
For exemple, if we consider the modulated initial state $S(x)=\cos(2\pi x/L)$, with $L\gg 1$,  we obtain 
in the long time regime $t\gg L$ 
\begin{equation}
m(x,t)=\cos(2\pi x/L) f_{\kappa}(2\pi t/L)\; ,
\end{equation}
that is the modulation does not spread with time but only the amplitude is decreasing as $t^{-1/2-\kappa}$.

Finally, for a homogeneous initial state with $S(x)=m(0)$, it is easy to see that $m(t)=m(0)$ in the XX case,
since the dynamic is conservative,  
and $m(t)=1/2m(0)$ for the Ising chain.

\section{Summary}
We have calculated for the critical Ising and $XX$ quantum chains the nonequilibrium transverse 
magnetization profiles for  kink-like and droplet-like initial states.
In both cases at large times the magnetization profiles exhibit scaling forms $t^{-1}F(l/t)$ which 
have been obtained analytically and are in excellent agreement with the numerics. 
The two systems show essentially the same features even if the 
dynamics of the transverse magnetization are very different, conservative for the $XX$ chain and nonconservative 
for the Ising model. In the Ising case there is a transient regime where the initial magnetization relaxes toward
the homogeneously initialy magnetized state stationary value $\pm 1/2$. After this initial regime, the system evolves 
as if the dynamics of the residual transverse magnetization was conservative. 
The long time relaxation of the transverse magnetization starting with a general initial $z$-state is expressed very simply
in terms of a convolution product of the initial distribution with a response kernel $f_{\kappa}$
obtained analytically for both XX and Ising chain.

\section*{Appendix: Time evolution }
The diagonalisation of the Hamiltonian (\ref{Hgam}) leads to~\cite{hinri}
\begin{equation}
H=i\sum_{q}\frac{\epsilon_{q}}{2}\gamma^{1}_{q}\gamma^{2}_{q}\; .
\end{equation}
In the Fermi operator representation, with $\eta_{q}=1/2(\gamma_{q}^{1}+i\gamma_{q}^{2})$ and $\eta_{q}^{\dagger}$ 
the hermitian conjugate,
one obtains $H=\sum_{q}\epsilon_{q}\eta_{q}^{\dagger}\eta_{q}-(1/2)\sum_{q}\epsilon_{q}$.
The time evolution of the Clifford operators is given by $U^{\dagger}_{q}(t)\gamma_{q} U_{q}(t)$ with
\begin{equation}
U_{q}(t)=\exp\left(\frac{\epsilon_{q}t}{2}\gamma_{q}^{1}\gamma_{q}^{2}\right)=\cos \frac{\epsilon_{q}t}{2}
+\gamma_{q}^{1}\gamma_{q}^{2}\sin \frac{\epsilon_{q}t}{2} \; ,
\end{equation}
which leads to equation~(\ref{rot}). Since $\{\gamma_{q}^{i},\gamma_{q'}^{j}\}=2\delta_{ij}\delta_{qq'}$, we can write equivalently
for equation~(\ref{rot}) $\gamma_{q}^{i}(t)=\sum_{j=1}^{2}\langle \gamma_{q}^{j}|\gamma_{q}^{i}(t)\rangle\gamma_{q}^{j}$, 
where the symbol $\langle.|.\rangle$
means the half of the anticommutator.

The time evolution of the $\Gamma$s is then expressed as 
\begin{equation}
\Gamma^{1}_{k}(t)= \sum_{q}\phi_{q}(k)\cos(\epsilon_{q}t) \gamma^{1}_{q}+\phi_{q}(k)\sin(\epsilon_{q}t)\gamma^{2}_{q}
\end{equation}
and
\begin{equation}
\Gamma^{2}_{k}(t)= \sum_{q}-\psi_{q}(k)\sin(\epsilon_{q}t) \gamma^{1}_{q}+\psi_{q}(k)\cos(\epsilon_{q}t)\gamma^{2}_{q}
\end{equation}
with initial values $\Gamma^{1}_{k}(0)=\sum_{q}\phi_{q}(k) \gamma^{1}_{q}$ and $\Gamma^{2}_{k}(0)=\sum_{q}\psi_{q}(k) \gamma^{2}_{q}$.
Reinjecting in this expressions the inverse transforms $\gamma^{1}_{q}=\sum_{k}\phi_{q}(k) \Gamma^{1}_{k}$ and 
$\gamma^{2}_{q}=\sum_{k}\psi_{q}(k) \Gamma^{2}_{k}$
one finally arrives at equations~(\ref{exp1}) with components~(\ref{comp}).

Formally, since the anticommutators $\{\Gamma_{k}^{i},\Gamma_{l}^{j}\}=2\delta_{ij}\delta_{kj}$ 
are all proportional to the identity operator,
the set $\{\Gamma_{k}^{i}\}$ forms an orthonormal basis of a $2L$-dimensional linear vector space $\cal E$ with 
inner product defined by $\langle.|.\rangle\equiv \frac{1}{2}\{.,.\}$. Hence, every vector $X\in {\cal E}$ has a unique 
expansion $X= \sum_{i,k}\langle \Gamma_{k}^{i}|X\rangle \Gamma_{k}^{i}$.
The string expression $X_{1}X_{2}...X_{n}$, with $X_{j} \in {\cal E}$, is a direct product vector of the space 
${\cal E}_{1}\otimes {\cal E}_{2}\otimes...\otimes{\cal E}_{n}$  which decomposition is
\begin{equation}
X_{1}X_{2}...X_{n}=\!\!\!\!\sum_{i_{1},k_{1},...,i_{n},k_{n}}
\langle \Gamma_{k_{1}}^{i_{1}}|X_{1}\rangle...\langle \Gamma_{k_{n}}^{i_{n}}|X_{n}\rangle 
\Gamma_{k_{1}}^{i_{1}}...\Gamma_{k_{n}}^{i_{n}}
\end{equation}
Using this formalism, the local magnetization at time $t$ is given by 
\begin{eqnarray}
\sigma_{l}^{z}(t)&=&-i\Gamma_{l}^{2}(t)\Gamma_{l}^{1}(t)\nonumber\\
&=&-i\sum_{i_{1},k_{1},i_{2},k_{2}}
\langle \Gamma_{k_{1}}^{i_{1}}|\Gamma_{l}^{2}(t)\rangle
\langle \Gamma_{k_{2}}^{i_{2}}|\Gamma_{l}^{1}(t)\rangle
\Gamma_{k_{1}}^{i_{1}}\Gamma_{k_{2}}^{i_{2}}
\end{eqnarray}
which is our starting point. One has then to consider the simple time-independant expectation values 
$\langle \Psi|\Gamma_{k_{1}}^{i_{1}}\Gamma_{k_{2}}^{i_{2}}| \Psi\rangle$, which are easilly obtained 
in the spin basis using the Jordan-Wigner expressions~(\ref{jw}).


\begin{thebibliography}{}
\bibitem{weiss} U. Weiss, {\it Quantum dissipative systems} (2nd edition) 
Series in Modern Condensed Matter Physics-Vol. 10, 1999 (World scientific Singapore)
\bibitem{racz1} T. Antal, Z. R\'acz and L. Sasv\'ari, Phys. Rev. Lett. {\bf 78}, 167 (1997)
\bibitem{racz2} T. Antal, Z. R\'acz, A. R\'akos and G. M. Sch\"utz, Phys. Rev. E  {\bf 59},  4912 (1999)
\bibitem{berim} G. O. Berim, S. Berim and G. G. Cabrera arXiv:cond-mat/0105594
\bibitem{trimper} G. M. Sch\"utz and S. Trimper, Europhys. Lett. {\bf 47}, 164 (1999)
\bibitem{igloi} F. Igl\'oi and H. Rieger, Phys. Rev. Lett. {\bf 85}, 3233 (2000)
\bibitem{lieb} E. Lieb, T. Schultz and D. Mattis, Ann. Phys. (N.Y.) {\bf 16}, 407 (1961)
\bibitem{hinri} H. Hinrichsen,  J. Phys. A: Math. Gen. {\bf 27}  5393 (1994)
\bibitem{turban} L. Turban and B. Berche, J. Phys. A: Math. Gen. {\bf 26}  3131 (1993)
\bibitem{remarq} The normalisation of the Hamiltonian is such that the sound velocity is one.
\bibitem{grad} I. S. Gradshteyn and I. M. Ryszhik, {\it Table of Integrals Series and Products} (Academic, London, 1980)
\end{thebibliography}
\end{document}